\documentclass{appolb}

\usepackage{amsmath,amssymb}

\begin{document}

\def\runhead{}

\title{
  Precision calculations in BR($\bar B\to X_s\gamma$)
  \thanks{
    Presented at
    ``Flavianet Topical Workshop, Low energy constraints on extensions of the Standard Model'',
    23-27 July 2009, Kazimierz, Poland
  }
}
\author{
  Thorsten Ewerth
  \address{
    Dipartimento di Fisica Teorica, Universit\`a degli Studi di Torino \& INFN Torino, I-10125 Torino, Italy
  }
}
\maketitle

\vskip 1cm

\begin{abstract}
  We briefly summarize the current status of perturbative calculations at next-to-next-to-leading-order (NNLO)
  accuracy in the $\bar B\to X_s\gamma$ decay rate as well as that of non-perturbative power-corrections.
\end{abstract}

\vskip 1cm

\section{Introduction}

Corrections to the $\bar B\to X_s\gamma$ decay are usually described in the framework of an effective
theory,\footnote{In writing (\ref{eq::lagrangian}) we discarded terms proportional to $V_{ub}V_{us}^*$ since they
  give only small contributions to the branching ratio that start at next-to-leading-order (NLO). Similar NNLO
  corrections can therefore be safely neglected.}
\begin{equation}\label{eq::lagrangian}
  {\cal L}_\text{eff} =
  {\cal L}_\text{QCD$\times$QED}(u,d,s,c,b) +
  \frac{4G_F}{\sqrt{2}}V_{tb}V_{ts}^*\sum_{i=1}^{8}C_i^{\rm eff}(\mu)O_i\,.
\end{equation}
Here, $C_i^{\rm eff}$ are renormalization scale dependent effective couplings, the so-called Wilson coefficients,
which encode the heavy gauge boson and the heavy top quark effects. The $b$-quark scale contributions, on the other
hand, are seen as matrix elements of flavor changing operators,
\begin{align}
 O_1 &= (\bar s_L\gamma_\mu T^a c_L)(\bar c_L\gamma^\mu T^a b_L)\,, &
 O_2 &= (\bar s_L\gamma_\mu c_L)(\bar c_L\gamma^\mu b_L)\,,\nonumber\\[2mm]
 O_{3,5} &= (\bar s_L\Gamma_{3,5} b_L){\sum}_q(\bar q\Gamma_{3,5}^\prime q)\,, &
 O_{4,6} &= (\bar s_L\Gamma_{3,5} T^a b_L){\sum}_q(\bar q\Gamma_{3,5}^\prime T^a q)\,,\nonumber\\[1mm]
 O_7 &= \frac{\alpha_{\rm em}}{4\pi}\,m_b(\bar s_L\sigma^{\mu\nu}b_R)F_{\mu\nu}\,, &
 O_8 &= \frac{\alpha_{s}}{4\pi}\,m_b(\bar s_L\sigma^{\mu\nu} T^a b_R)G^a_{\mu\nu}\,,
\end{align}
where $\Gamma_3=\gamma_\mu$, $\Gamma_3^\prime=\gamma^\mu$, $\Gamma_5=\gamma_\mu\gamma_\nu\gamma_\lambda$ and
$\Gamma_5^\prime=\gamma^\mu\gamma^\nu\gamma^\lambda$. Using (\ref{eq::lagrangian}), the differential decay rate for
$\bar B\to X_s\gamma$ can be written as follows,
\begin{equation}
  d\Gamma =
  \frac{G_F^2\alpha_{\rm em}m_b^2}{256\pi^6m_B}\,
  |V_{tb}V_{ts}^*|^2\,
  \frac{d^3q}{E_\gamma}
  \sum_{i,j}C_i^{\rm eff}(\mu)C_j^{\rm eff}(\mu)W_{ij}(\mu)\,.
\end{equation}
In the equation displayed above, $q$ denotes the momentum of the photon and the $W_{ij}$, which describe the
hadronic dynamics, are given by the imaginary part of forward scattering amplitudes,
\begin{equation}\label{eq::wij}
  W_{ij}(\mu) =
  2\,{\rm Im}\left(
  i\int\!d^4x\,e^{-iq\cdot x}\langle\bar B|\,T\!\left\{O_i^\dagger(x)O_j(0)\right\}|\bar B\rangle
  \right)\,.
\end{equation}
Since the mass of the $b$-quark is much larger than the binding energy of the $B$-meson, which is of the order
of $\Lambda\equiv\Lambda_{\rm QCD}$, we can perform an operator product expansion (OPE) of the time ordered products.
Doing so, one finds that the leading term is the partonic decay rate which gives the dominant contribution, while
the non-leading terms, the so-called power-corrections, are suppressed by powers of $\Lambda/m_b$ and give
non-vanishing contributions starting from $O(\Lambda^2/m_b^2)$.\footnote{Strictly speaking, equation
  (\ref{eq::wij}) and its OPE hold only for $W_{77}$. In all other cases the $W_{ij}$ contain contributions in which
  the photon couples to light quarks ($u,d,s,c$), and this leads to non-perturbative effects different from that
  mentioned above (see section \ref{sec::non-perturbative}). Furthermore, for $W_{ij}\not=W_{77}$, contributions
  that contain no photon in the final state should be excluded, as required by the experimental setup.} In what
follows we describe the state-of-the-art of perturbative and non-perturbative corrections in the
$\bar B\to X_s\gamma$ decay.

\section{Perturbative corrections}

The calculation of the perturbative corrections can be divided into three steps. In the first step one has to
evaluate the effective couplings $C_i^{\rm eff}$ at the high-energy scale $\mu\sim M_W$ by requiring equality of
the Standard Model and the effective theory Green functions. Defining $\tilde\alpha_s(\mu)=\alpha_s(\mu)/(4\pi)$,
the effective couplings can be expanded as follows,
\begin{equation}
  C_{i}^{\rm eff}(\mu) =
  C_{i}^{(0)\rm eff}(\mu) +
  \tilde\alpha_s(\mu)C_{i}^{(1)\rm eff}(\mu) +
  \tilde\alpha_s^2(\mu)C_{i}^{(2)\rm eff}(\mu) +
  \dots\,.
\end{equation}
At NNLO accuracy one has to determine the coefficients $C_{i}^{(2)\rm eff}(\mu)$. For $i\,$=$\,7,8$ it required
performing a three-loop calculation \cite{Misiak:2004ew} whereas for the remaining cases $i\,$=$\,1,\dots,6$ a
two-loop calculation was sufficient \cite{Bobeth:1999mk}.

The second step involves the calculation of the anomalous dimension matrix $\gamma^{\rm eff}$ which describes
the mixing of the operators under renormalization. Its knowledge is necessary to solve the effective theory
renormalization group equations for the effective couplings,
\begin{equation}
  \mu\frac{d}{d\mu}C_i^{\rm eff}(\mu) = \sum_j\gamma_{ji}^{\rm eff}C_j^{\rm eff}(\mu)\,,
\end{equation}
and to evolve the latter down to the low-energy scale $\mu\sim m_b$. Performing a perturbative expansion in
the strong coupling constant, the anomalous dimension matrix takes the following form,
\begin{equation}
  \gamma^{\rm eff} =
  \tilde\alpha_s(\mu)\gamma^{(0)\rm eff} +
  \tilde\alpha_s^2(\mu)\gamma^{(1)\rm eff} +
  \tilde\alpha_s^3(\mu)\gamma^{(2)\rm eff} +
  \dots\,.
\end{equation}
At NNLO one has to determine $\gamma^{(2)\rm eff}$ which is actually a $8\times8$ matrix,
\begin{equation}
  \gamma^{(2)\rm eff} =
  \begin{pmatrix}
    A_{6\times 6}^{(2)} & B_{6\times 2}^{(2)}\\[1mm]
    0_{2\times 6} & C_{2\times2}^{(2)}
  \end{pmatrix}\,.
\end{equation}
The block matrices $A$ and $C$ describing the self-mixing of the four-quark operators and the self-mixing of the
dipole operators at three loops, respectively, have been calculated in \cite{Gorbahn:2004my}. The block matrix $B$
describing the mixing of the four-quark operators into the dipole operators at four loops has been determined in
\cite{Czakon:2006ss}. After this calculation the first two steps of the perturbative calculation were completed,
that is the effective couplings at the low-energy scale $\mu\sim m_b$ with resummed logarithms are now known at
NNLO accuracy.\footnote{This means large logarithms have been resummed up to $O(\alpha_s^{n+2}\ln^n(m_b/M_W))$.}

In the last step one has to calculate on-shell amplitudes of the operators at the low-energy scale. This is the
most difficult part of the NNLO enterprise and it is still under investigation. In order to see what has been done
so far, and what still has to be done, we write the decay rate for the partonic decay
$b\to X_s^{\rm partonic}\gamma$ as follows,
\begin{equation}\label{eq::gamma-partonic}
  \left.\Gamma^{\rm partonic}\right|_{E_\gamma>E_0} =
  \frac{G_F^2\alpha_{\rm em}m_b^5}{32\pi^4}\,
  |V_{tb}V_{ts}^*|^2\,
  \sum_{i,j}C_i^{\rm eff}(\mu)C_j^{\rm eff}(\mu)G_{ij}(E_0,\mu)\,,
\end{equation}
where $G_{ij}(E_0,\mu)$ can again be expanded in terms of $\tilde\alpha_s$,
\begin{equation}
  G_{ij}(E_0,\mu) =
  Y_{ij}^{(0)}\delta_{i7}\delta_{j7} +
  \tilde\alpha_s(\mu)Y_{ij}^{(1)}(E_0,\mu) +
  \tilde\alpha_s^2(\mu)Y_{ij}^{(2)}(E_0,\mu) +
  \dots\,.
\end{equation}
At NNLO one has to determine the coefficients of $\tilde\alpha_s^2(\mu)$ which, however, has only been done
in a complete manner for $i\,$=$\,j\,$=$\,7$ \cite{Blokland:2005uk,Asatrian:2006rq}. Once we neglect on-shell
amplitudes that are proportional to the small Wilson coefficients of the four-quark penguin operators $O_3$-$O_6$,
the remaining cases to be considered are $(ij)\,$=$\,(11),\,(12),\,(22),\,(17),\,(18),\,(27),\,(28),\,(78)$,
and $(88)$. The large-$\beta_0$ corrections are known in all these cases except for $(18)$ and $(28)$
\cite{Bieri:2003ue,Ligeti:1999ea,88}. In addition, effects of the charm and bottom quark masses on the gluon lines
are known in all the cases \cite{Ewerth:2008nv,Boughezal:2007ny}. The other beyond-large-$\beta_0$ corrections
have been found only in the limit $m_c\gg m_b/2$, except for the $(78)$ and $(88)$ cases \cite{Misiak:2006ab}. This
limit has been used to interpolate the unknown beyond-large-$\beta_0$ corrections at $O(\alpha_s^2)$ to the measured
value of $m_c\approx m_b/4$ \cite{Misiak:2006ab}. The result for the branching ratio, for $E_0=1.6\,{\rm GeV}$, is
given by \cite{Misiak:2006zs}\footnote{For a discussion of the residual renormalization scale dependence of the
  branching ratio at NNLO we refer the reader to \cite{Misiak:2006zs}.}
\begin{equation}\label{eq::nnlo-estimate}
  \text{BR}(\bar B\to X_s\gamma)_{\rm SM} =
  (3.15\pm 0.23)\times 10^{-4}\quad (\mbox{1S scheme})\,.
\end{equation}
The theoretical uncertainty of this NNLO estimate is at the same level as the uncertainty of the current world
average reported by HFAG \cite{Barberio:2008fa},\footnote{This average includes the measurements from CLEO and
  BaBar and Belle \cite{Chen:2001fja}. The recently published update by Belle \cite{Limosani:2009qg} has not
  been taken into account.}
\begin{equation}\label{eq::hfag}
  \text{BR}(\bar B\to X_s\gamma)_{\rm exp} =
  \left(3.52\pm 0.23\pm 0.09\right)\times 10^{-4}\,,
\end{equation}
which is furthermore expected to come down to the 5\% level at the end of the B-factory era.

Here a remark concerning the overall normalization of the theoretical prediction is in order. To reduce parametric
uncertainties stemming from the CKM angles as well as from the $c$- and $b$-quark masses, the partonic decay rate
given in (\ref{eq::gamma-partonic}) is usually normalized using a combination of the $\bar B\to X_cl\bar\nu$ and
$\bar B\to X_ul\bar\nu$ decay rates which is reflected by the appearance of the semileptonic phase-space factor
\begin{equation}
  C = \left|\frac{V_{ub}}{V_{cb}}\right|^2
  \frac{\Gamma(\bar B\to X_cl\bar\nu)}{\Gamma(\bar B\to X_ul\bar\nu)}
\end{equation}
in the analytical expressions \cite{Gambino:2001ew}.\footnote{The
  denominator $\Gamma(b\to ul\bar\nu)$ is already known at NNLO accuracy \cite{vanRitbergen:1999gs}.}
 Unfortunately, the
determination of $m_c$ and $C$ from the fit from the measured spectrum of the $\bar B\to X_cl\bar\nu$ decay
in the 1S scheme \cite{Bauer:2004ve} differs from that in the kinetic scheme \cite{Gambino:2008fj}.\footnote{In
  \cite{Misiak:2006zs} the values for $m_c$ and $C$ from \cite{Bauer:2004ve} were adopted.} Using the values for
$m_c$ and $C$ of the latter determination results in a higher central value for the $\bar B\to X_s\gamma$ decay
rate \cite{Misiak:2008ss},
\begin{equation}
  \text{BR}(\bar B\to X_s\gamma)_{\rm SM} =
  \left(3.25\pm 0.24\right)\times 10^{-4}\quad (\mbox{kinetic scheme})\,.
\end{equation}
The difference of $m_c$ and $C$ in the 1S and kinetic scheme is likely to be due to different input data,
differences in the fit method, and treatment of theory errors. In this respect, supplementing the fit, for example,
by the determination of the $c$- and $b$-quark masses from sumrules \cite{Chetyrkin:2009fv} could possibly be
helpful to reduce the discrepancy of $C$ in both schemes.

We should also remark that not all of the aforementioned contributions to the function $G_{ij}$ entered the
analysis of \cite{Misiak:2006zs}. These are the massive fermionic corrections presented in
\cite{Asatrian:2006rq,Ewerth:2008nv,Boughezal:2007ny} and the large-$\beta_0$ contributions for
$(ij)\,$$\not=$$\,(77)$ from \cite{Ligeti:1999ea,88}.\footnote{Also the mixing of the four-quark operators
  $O_{1-6}$ into the chromomagnetic dipole operator $O_8$ \cite{Czakon:2006ss} was not included in
  \cite{Misiak:2006zs}.} These contributions will be included in a future update together with so far unknown
contributions, which is, for example, the complete knowledge of $G_{78}$ \cite{78}. Also the complete
calculation of $G_{27}$ for $m_c=0$ is underway \cite{Schutzmeier:2008sm}. Especially the latter will prove useful
in the reduction of the uncertainty stemming from the interpolation in $m_c$. Apart from the NNLO corrections also
tree-level diagrams with the $u$-quark analogues of $O_{1,2}$ and the four-quark operators $O_{3-6}$ have been
neglected so far \cite{treelevel}. The numerical effect of all of these contributions on the branching ratio is or
is expected to remain within the uncertainty of the NNLO estimate given in (\ref{eq::nnlo-estimate}).

Finally, we should mention that there are also cutoff-enhanced corrections which matter close to the endpoint
\cite{Becher:2005pd}. However, as demonstrated in \cite{Misiak:2008ss}, the resummation of the cutoff-enhanced
logarithms overestimates the effect of the $O(\alpha_s^3)$-terms for $E_0\lesssim 1.6\,{\rm GeV}$. Therefore, the
prediction for the branching ratio given in (\ref{eq::nnlo-estimate}) should, at present, be considered as
more reliable.

\section{Non-perturbative corrections}
\label{sec::non-perturbative}

Since the perturbative calculations in $\bar B\to X_s\gamma$ are now performed at NNLO the non-perturbative
corrections become more important. In general well under control are the power-corrections stemming from the
OPE of the time ordered products contained in (\ref{eq::wij}). For the self-interference of the electromagnetic
operator, that is for $i\,$=$\,j\,$=$\,7$, they are known at $O(\Lambda^2/m_b^2)$ \cite{Falk:1993dh} and
$O(\Lambda^3/m_b^3)$ \cite{Bauer:1997fe}. For all other combinations of $i$ and $j$ the time ordered products
include contributions in which the photon couples to light quarks, and that causes the breakdown of the OPE. In
this case non-perturbative collinear effects \cite{Kapustin:1995fk} as well as power-corrections at
$O(\Lambda^2/m_c^2)$ show up \cite{Voloshin:1996gw}. The combined effect of all of the aforementioned
non-perturbative corrections is of around 3\% in the branching ratio. Besides, non-perturbative effects appearing
at $O(\alpha_s\Lambda/m_b)$ show up when the photon couples to light quarks. Their size is not known at present
and hence a 5\% uncertainty related to all the unknown non-perturbative effects has been included in
(\ref{eq::nnlo-estimate}). The size of this uncertainty is supported by the estimate of the
$O(\alpha_s\Lambda/m_b)$-corrections in the interference of the electro- and chromomagnetic dipole operators
performed in \cite{Lee:2006wn}. As pointed out in references \cite{Lee:2006wn,mikolaj}, the magnitude
of the effect considered in \cite{Lee:2006wn} could be probed by an improved measurement
\cite{Aubert:2005cua} of the isospin asymmetry
\begin{equation}
  \Delta_{0-} =
  \frac{\Gamma(\bar B^0\to X_s\gamma)-\Gamma(\bar B^-\to X_s\gamma)}
       {\Gamma(\bar B^0\to X_s\gamma)+\Gamma(\bar B^-\to X_s\gamma)}\,.
\end{equation}
Finally, we note that a non-perturbative uncertainty appears also when extrapolating the three different
measurements performed at CLEO, BaBar and Belle down to the common lower cut $E_0=1.6\,{\rm GeV}$ in the photon
energy. It is accounted for in the error of the world average given in (\ref{eq::hfag}).

\section{Conclusions}

At present the uncertainties in the branching ratio of $\bar B\to X_s\gamma$ are on the same level on both
the theoretical and experimental side. Thanks to the ongoing calculations of the perturbative corrections the
uncertainty stemming from this part will further reduce. However, to reach the 5\% level or even less on the
theoretical side a better understanding of the non-perturbative power-corrections at $O(\alpha_s\Lambda/m_b)$
is required.

\vskip .5cm


\noindent {\it Acknowledgements:} I am grateful to H.~Czyz, M.~Krawczyk, and M.~Misiak for the invitation to
this workshop. I also would like to thank P.~Gambino, and especially M.~Misiak, for comments on the final version
of this manuscript. This work was supported by MIUR under contract 2004021808-009 and by a European Community's
Marie-Curie Research Training Network under contract MRTN-CT-2006-035505 `Tools and Precision Calculations for
Physics Discoveries at Colliders'.

\end{document}